Species diversity in a metacommunity with patches connected by periodic coalescence: a neutral model.

Janis Antonovics[1], Stavros D. Veresoglou[2,3,4], Matthias C. Rillig[2,3,5]

[1] Biology Department, University of Virginia, Charlottesville, VA22904, USA
[2] Institut für Biologie, Freie Universität Berlin, 14195 Berlin, Germany.
[3] Berlin-Brandenburg Institute of Advanced Biodiversity Research (BBIB), 14195 Berlin, Germany.

[1] Corresponding author. e-mail: ja8n@virginia.edu, ORCID: 0000-0002-5189-1611

*Mailing address*: Biology Department, University at Virginia, 485 McCormick Road, P.O. Box 400328, Charlottesville, VA 22904, USA. *Phone*: 1-434-243-5077

[4] e-mail: veresoglou@gmail.com

[5] e-mail: rillig@zedat.fu-berlin.de

**Abstract**

The recent realization that entire communities fuse and separate (community coalescence) has led to a reappraisal of the forces determining species diversity and dynamics, especially in microbial communities where coalescence is likely widespread. To understand if connectedness by coalescence results in different outcomes from connectedness by individual dispersal, we investigated chance processes leading to loss of species diversity using a model of a neutral two-species metacommunity. Two scenarios were investigated: (a) 'pairwise coalescence' where the communities coalesce in pairs, intermix and then separate; (b) 'diffuse coalescence' where several communities mix as a pool and are re-distributed to their original patches. When standardized for the same net movement, both types of coalescence led to a longer time to single species dominance than dispersal. Coalescence therefore may be an important process contributing to the surprisingly high microbial species diversity in nature.

**Introduction**

    The realization that local communities are not independent but interact with each other through movement of individuals (Wilson 1992; Leibold et al. 2004) has greatly increased our understanding of the drivers of local and regional species diversity (Matthiessen & Hillebrand 2006; Venail et al. 2008). It has been proposed that in microbial systems another process, namely community coalescence, may be especially important (Rillig et al. 2015). In this process the whole community (perhaps together with the habitat) may wholly or partly merge with another community. This might occur, for example, when an endophyte community in a leaf contacts the soil community during leaf fall, when microbiomes are transferred during close human oral or genital contact, or when a group of ponds are transiently connected by periodic flooding or by tides. There is compelling evidence that coalescent events happen in nature in both terrestrial and aquatic habitats, especially at the microbial scale (Rillig et al. 2016; Mansour et al. 2018). Many factors may influence the outcome of coalescence, including the frequency of the coalescence events, the level and type of concurrent environmental exchange, and whether the coalescence is wholesale or consists of the exchange of only a small fraction of the original communities (Rillig et al. 2015).

    A recent theoretical study (Tikhonov 2016), showed that communities even in the absence of direct cooperation between the interacting species tended to persist following coalescence, indicating community level cohesion, as first suggested by Gilpin (1994). This theoretical possibility has been confirmed by experimental studies showing that anaerobic communities with the greatest productivity as measured by methane production, were the ones that tended to persist following coalescence (Sierocinski et al. 2017).

    To develop a theoretical framework for understanding how coalescence events affect metacommunity dynamics, we compare a neutral model of communities connected by coalescence with a neutral model of communities connected by individual dispersal, yet with equivalent levels of overall movement. We investigate two types of coalescent process. In one type, there is pairwise contacts among patches, their communities merge, and then separate. The biology here can be envisaged to occur, for example, with oral or genital contacts in humans where the coalescence is transient, pairwise and relatively symmetrical. Such pairwise coalescence could be called the "kiss-and-run" model. In the other type of coalescence, which we term diffuse, we envisage a group of ponds or rock-pools that are periodically flooded, where

ponds merge, but then as the flood recedes, the component species recolonize those same pools at random. Such diffuse coalescence could be called the "flooded-ponds" model. In many ways, these processes are analogous to the "stepping-stone" and "island" models of dispersal in population genetics (Wright 1943; Kimura 1953).

Our results show that time to monodominance is generally longer for communities connected by coalescence than by dispersal, which implies that if much of the dynamics is transient, coalescence may be an important factor contributing to the observed high species diversity of many microbial systems.

## Methods

Throughout, we assume a metacommunity of $N$ individuals belonging to two species distributed over a set of $p$ patches (or islands) each consisting of $n$ individuals ($N = pn$). The model is neutral in the sense that individuals of both species have the same birth, death, and movement rates, but these processes are instantiated stochastically (see below). We assume patch structure is deterministic ($p$ and $n$ are constant). Movement among patches occurs either by dispersal of single individuals or by patches becoming interconnected (coalescing). Throughout, we use the term *dispersal* in a technical sense to represent movement of *single individuals* among patches (the classical metacommunity scenario), and the term *coalescence* in a technical sense to represent movement that results from *fusion of communities* that are in the patches. Coalescence refers to *community* coalescence *sensu* Rillig et al. (2016) and not to coalescence through ancestry as in population genetics. We use the term *movement* to refer to the transfer of individuals among patches of the metacommunity regardless of the mechanism by which this is achieved. We assume movement does not change patch size. Coalescence and dispersal represent complementary mechanisms of movement in the metacommunity, and microbial systems most likely experience both. To disentangle their relative effects, we assume the patches are connected exclusively either by coalescence events or by dispersal events.

As the common unit of movement, we used the parameter $m$, representing the probability that an individual will be found in a different patch from its original location after a single time step. Note that this is not the per patch movement rate, as this latter will vary with the degree of subdivision of the metacommunity. We used two implementations of stochastic birth-death processes known as the Moran and the Fisher-Wright models (Blythe and McKane 2007). The Fisher-Wright model permitted deterministic implementation of movement, so ensuring differences between dispersal and coalescence were not due to the way stochasticity was implemented for the two types of movement.

Usually, when these two model structures are implemented in systems with patch structure, movement occurs as part of the birth-death process with dispersal being instantiated by assuming that a proportion of the individuals that die in a patch are replaced by births from other patches (Moran 1959, 1962; Wakeley & Takahashi 2004). However, to accommodate dispersal and coalescence equivalently, we separated the stochastic within-patch birth-death processes from the movement processes (Parra-Rojas and McKane 2018). There are strong parallels between community subdivision in ecology and population subdivision in population genetics, and that the model structure presented here applies equally to two haploid genotypes within a metapopulation (Blythe & McKane 2007). Coalescent scenarios are therefore also generalizable to genetic variants within species. For example, in social animals, whole groups may coalesce and separate such that changes in the organizational structure of social groups modifies allele frequencies (Mihaljevic 2012).

*Moran model*

*Birth-death:* At each time step, an individual in the metacommunity died and was replaced by an individual chosen randomly based on its frequency in the patch prior to the death (Moran 1958, Hubbell 1979). This metacommunity was therefore composed of individuals with overlapping generations, each with an expected life span of $1/N$ time steps. Normally, this Moran process is applied to a single species or patch, but because we wanted to assess the effects of subdivision for comparisons between coalescence and dispersal, $N$ here is the number of individuals in the metacommunity.

*Dispersal:* To equate as far as possible the stochastic processes occurring in dispersal and coalescence, we simulated dispersal by a process analogous to the way we instantiated pairwise coalescence (see below). At each time step, with probability $m$, we identified one random individual in each of two different patches. These individuals were then re-assigned to their source patches at random with equal probability. In this way, half the re-assignments resulted in no movement events, whereas half of them resulted in the movement (exchange) of two individuals, thus averaging one individual being dispersed between patches. No more than one dispersal event per metacommunity per time step was allowed.

*Pairwise coalescence*: For pairwise coalescence (the "kiss-and-run" model), at each time step following birth-death, a 'focal' patch was chosen at random with probability $c=m/n$ and the individuals in that patch were pooled with those in another patch. Individuals were then randomly re-assigned into the two patches, such that on average half the individuals returned to their original patch while half moved between patches. Therefore, one coalescence event between two patches of *n* individuals resulted in an average of *n* movement events. No more than one coalescent event per metacommunity per time step was allowed. In effect, dispersal is frequent movement of a few individuals among patches, whereas pairwise coalescence events are less frequent, but each event moves more individuals between fewer patches.

*Diffuse coalescence*: More movement results from patches coalescing diffusely than from the same number of patches coalescing pairwise. Thus, for example, if four patches of *n* individuals are involved in pairwise coalescence, the total amount of movement is $4n/2$ (each coalescent event on average moves half the individuals in each patch). However, if four patches of *n* individuals are involved in diffuse coalescence, then the probability of any individual returning to its patch of origin is 0.25 and total movement is $4n*(1-0.25)$. Generalizing, in *p* patches of size *n*, movement by pairwise coalescence is proportional to *pn/2*, while movement by diffuse coalescence is *pn(1-(1/p))*; as the number of patches involved in coalescence increases, there will be proportionately more net movement if the coalescence is diffuse than if it is pairwise (see Fig. Supplementary material for an explanatory diagram). Because diffuse coalescence moves more individuals than pairwise coalescence, for any level of diffuse coalesence involving a specific number of patches, we instantiated that event proportionately less frequently. For example, coalescence of 4 patches of size *n* moves $3n$ individuals, so we instantiated diffuse coalescence three time less frequently than pairwise coalesence (i.e. with probability $m/3n$).

*Differential reproductive output*: We also briefly investigated if the difference between coalesence and dispersal would change in a simple non-neutral case, namely when there was a differential advantage of one species over another among genotypes). During the Moran birth death process, we instantiated one type to have an advantage, *s*, by adjusting the probability that it would replace the individual that dies to: $(1+s)f_i/((1+s)f_i+(1-f_i))$, where $f_i$ is the frequency of the advantageous type in patch *i*.

*Fisher-Wright model*

We used the Fisher-Wright model to instantiate dispersal and coalescence as a deterministic process, so that differences in stochastic implementation of the two movement types would not be a confounding factor. To do this, for any metacommunity size, we chose patch sizes that made it possible to equate dispersal and coalescence in terms of the number of individuals moved per generation. As a numerical example, if the metacommunity consists of $N=128$ individuals, and the number of patches is $p=16$ then it is possible to move 8 individuals per generation ($m=8/128$) by either one coalescence event (in which half the individuals are exchanged between two patches) or by four dispersal events involving direct exchange of individuals among eight patches (without replacement). However, such equivalency, is only possible for some patch structures.

*Birth-death*: In this model, at each time interval, all individuals die and are replaced by the same number of individuals sampled at the same frequency as those present in the previous generation (Wright 1943), as if progeny were produced clonally. Individuals therefore have non-overlapping generations, and each time interval is a generation.

*Dispersal*: If as above, $m$ is the per individual probability of movement in the metacommunity, $m*N$ individuals move per time step. Single individuals were identified in this number of patches without replacement, the patches were paired, and the individuals exchanged.

*Pairwise coalescence*: Dispersal of $m*N$ individuals per generation is equivalent to $C=m*N/n$ coalescent events. If $C>p=2$, $2C$ patches were chosen without replacement, paired, and exactly half the individuals within each patch (chosen at random with regard to type) were assigned back to their two source patches.

*Diffuse coalescence:* To compare the effects of diffuse vs. pairwise coalescence, we used the following equivalency: net movement is the same when $p$ patches are involved in pairwise coalescence and when $(p/2)+1$ patches are involved in diffuse coalescence (see derivation above, for Moran model).

*General implementation*

We used computer simulation but confirmed components of our models by testing agreement with existing theory of fixation rates in subdivided populations (Kimura and Ohta 1968). All simulations were started with equal expected numbers of the two species in each patch (i.e. chosen with a 0.5 binomial expectation). Throughout, we present the results for $N=128$, and $p=2, 4, 8, 16, 32$ and $64$; these values resulted in reasonable run-times and permitted modeling equivalent levels of movement by dispersal and coalescence; other values of *N*, *p*, and *n* always gave qualitatively similar results. All programs used R version 3.5 and the default random number generator internal to the functions 'sample', 'runif', and 'rbinom'.

**Results**

When coalescence is pairwise, the time to monodominance (fixation) for any given rate of movement was longer when the patches were connected by coalescence than when they were connected by dispersal. This was seen under both the Moran and the Fisher Wright birth-death processes (Figs 1). The results were qualitatively same if the median of the time to dominance was considered (see Appendix S1 in Supporting Information, Table S1.1). In both models, the mean time to coalescence with no subdivision was as expected theoretically (Supplementary material B Text 2). In the Fisher-Wright model with deterministic movement, the time to monodominance at the highest level of community subdivision, i.e. *p*=64 and *n*=2, was the same for dispersal and coalescence (Fig. 2). This was because at *n*=2, our implementations of dispersal and coalescence are identical (see Appendix S2.1 in Supporting Information). Assigning individuals randomly to patches after coalescence rather than deterministically made little difference to the overall pattern (compare Fig. 2 and Fig. S1.1 in Appendix S1 in Supporting Information).

As expected, time to monodominance was always greater when there was more patch sub-division (numbers in each patch were smaller and random effects greater) and when the rates of movement among patches were lower (local chance divergence was greater among patches). Also as expected, because generations are not overlapping, the Moran model needed more time steps than the Fisher-Wright model to achieve monodominance.

When coalescence was diffuse, the time to monodominance was longer than when it was pairwise (Fig. 3). It is important to emphasize that these processes were standardized for total amount of movement, and not total number of patches involved. If the same number of patches are involved, then coalescing these diffusely rather than pairwise results in more effective movement and therefore shorter times to monodominance especially at high movement rates (Fig. S1.2 in Appendix S1 in Supporting Information).

The general result that the time to monodominance is longer under coalescence was unchanged by the addition of differences in reproductive rates of the two species (Fig 4a). As expected, time to monodominance of the favored species was shorter as its reproductive advantage increased, and not surprisingly the movement mode did not change the probability that the fitter type predominated (Fig. 4b) because movement was independent of the birth-death process.

## Discussion

Our results show that the neutral expectations of metacommunity subdivision on species abundance are different depending on whether the component communities are connected by dispersal of single individuals or by coalescence of whole communities, even when the total number of individuals moving by these routes is the same. Where coalesence is by fusion and separation of patches, the time to monodominance is longer, implying that if there were only neutral forces involved, the transient state where there is species diversity would be maintained for a longer time under coalescence than under dispersal. By implementing this process using the Fisher-Wright model with deterministic movement, we ensured that the results were due to the movement modes, and not to differences in the way stochasticity was implemented in the code used to simulate the two movement types.

The reason for the greater time to monodominance with coalescence is not obvious, and we undertook this study because we could not ourselves intuitively decide which outcome to expect. However, it likely that coalescence averages the differences among patches diverging by stochastic birth-death processes more effectively than dispersal. For monodominance at the metacommunity level, all patches must eventually end up with the *same* species. Two opposing forces determine this. First, monodominance occurs faster if all patches (or the average of those patches) are at starting frequencies greater or less than 0.5. In our simulations, all communities were started at an average frequency of 0.5 of the two species, so this was not a factor. Secondly, and somewhat counterintuitively, monodominance is more likely if the initial frequencies among the patches are different. For a species starting at frequency *p*, the time to reach monodominance is given by a function of the following form (Kimura & Ohta 1969):

$$- ((1-p)/p) \log(1-p)$$

In a metacommunity with just two patches with initial frequencies, $p_1$ and $p_2$, it can be shown that the average time to monodominance of a particular species is longer if the starting frequencies are the same ($p_1 = p_2 = \bar{p}$) than if they are different ( $p_1 \neq p_2$, and $(p_1 + p_2)/2 = \bar{p}$ ) (Fig. S1.3 in Appendix S1 in Supporting Information). Dispersal events would create small perturbations from existing frequencies, but pairwise coalescence events would equalize frequencies in separate patches that have diverged due to demographic stochasticity. This is consistent with the difference between time to monodominance for coalescence and for dispersal being greater when overall movement rates are lower (Fig. 1, 2).

It is also consistent with the times to monodominance being longer for diffuse coalescence. Diffuse movement is expected to drive frequencies in patches more towards the global average than pairwise coalescence, hence reduce the variance among patches, and correspondingly lead to longer times to monodominance. Our finding that diffuse coalescence leads to longer times to monodominance than pairwise carries the important qualification that this is true only if net movement by the two processes in the metacommunity is held the same. If different levels of coalescence are expressed simply as number of patches involved, then the more patches are involved the faster will be the time to monodominance because net movement rate is also greater: in the limit, if all patches coalesce in a highly subdivided community, it essentially becomes one well-mixed community.

In ecological models of stochastic processes, the Moran model has been most commonly used, whereas in genetic models the Fisher-Wright model has been preferred (Kimura & Ohta 1971). However, stochastic processes represented by both models are analogous, except in the Moran model individuals are effectively perennial, generations overlap, and the alternative states are usually conceptualized as representing species instead of alleles (Blythe & McKane 2007). When only dispersal (and not coalescence) events are invoked, as is usually done in models of community or population subdivision, the results for the two types of model are similar and reconcilable by rescaling generation time under the Moran model by a factor of $N/2$ relative to the Fisher-Wright model (Blythe & McKane 2007). It is therefore not surprising that the outcomes from the two types of model with regard to coalescence are also qualitatively similar. Unlike in most studies (but see Parra-Rojas & McKane 2018), we implemented the Moran and Fisher-Wright birth and death processes separately from the movement events, and this enabled us to confirm that the greater time to monodominance was due to the "batch" nature of the movement rather than because of differences in stochasticity.

The goal of the present study has been to investigate neutral processes, rather than the outcome of differential fitness of the components. Nevertheless, using a simple modification of the Moran process, to impose, we showed that simple differences in reproductive rates of the two species did not change the relative effects of coalescence vs. dispersal. However, it would be intriguing to explore more interesting and complex forms of selection, as coalescence transfers groups of individuals, and processes involving group interactions would likely differ between coalescence and dispersal. Such 'group interactions' might be processes involving altruism or shared 'public goods' (Archetti & Scheuring 2010), or positive density-dependent effects such as 'quorum sensing' (Perez-Velazquez et al. 2016). Coalescence may also affect interactants that show 'community cohesion' without any form of altruism (Tikhonov 2016). A second additional feature of coalescence that we haven't explored is that coalescence merges both abiotic and biotic aspects of the 'patches', and not necessarily to the same degree Rillig et al. 2016). It would be of interest to assess how the coalescence of, say, resource pools in patches with different species composition might affect species coexistence in a metacommunity.

The model presented here is heuristic and was not intended to apply to any specific system in nature. Our models were simplified, in that they assumed constant and equal patch-structure and uniform patterns of movement. Moreover, in any system with spatial sub-structuring, there is also likely to be extinction and recolonization of entire patches (Leibold & Loeuille 2015; Fukumori et al. 2015). Colonization and extinction can also be a major factor determining genetic structure of populations (Wade & McCauley 1988). Our results present an additional scenario, and therefore reinforce the need for caution in presuming any particular movement structure when interpreting data on species diversity (or on genetic differentiation) in subdivided habitats. The rapid evolution of the metacommunity concept over the last two decades and its entry into mainstream ecology (Hubbell 2001; Rosindell et al. 2011; Mihaljevic 2012; Leibold et al. 2014) reflects awareness that adding reality in ecological theory results in more accurate assessment of the forces determining species diversity (Matthiessen & Hillebrand 2006; Venail et al. 2008). In terrestrial systems, where ecological studies have been largely focused on 'large' organisms, mainly plants and animals, a theory based more on the dispersal of individuals rather than whole communities is clearly appropriate, but in microbial systems coalescence is likely to be an important additional element of realism.

There has been substantial advance in recent years in the development of analytical approaches to understanding stochastic processes especially in subdivided populations (Constable & McKane 2014, 2018; Parras-Rojas & McKane 2018) and we hope that our results will stimulate extended analytical efforts in the area of community coalescence. Microbial communities have the advantage over communities of macro-organisms in that it may be feasible to manipulate the frequency of different kinds of movement events experimentally and to examine their consequences in real organisms (Jessup et al. 2005; Fukumori et al. 2015). Translating these ideas to better understanding the diversity of natural microbial systems, where metacommunity dynamics is likely to be highly variable, is an important challenge.

## Acknowledgements

Support for this work came from the following: NIH Grant R01 GM122061 and a Humboldt Research Award to JA; DFG BioMove Research Training Group (DFG-GRK 2118/1) and ERC-AdG 'Gradual Change' (694368) to MCR. We are grateful to Alan McKane who provided helpful comments, and to Emme Bruns and Henry Wilbur for improving the manuscript.

## References

Archetti M, Scheuring I. 2010. Coexistence and cooperation and defection in public goods games. *Evolution* 65: 1140-1148.

Constable GWA, McKane AJ. 2014. Population genetics on islands connected by an arbitrary network: an analytical approach. *J Theoret Biol* 358: 149-165.

Constable GWA, McKane AJ. 2018. Exploiting fast variables to understand population dynamics and evolution. *J Stat Phys* 172: 3-43.

Blythe RA, McKane AJ. 2007. Stochastic models of evolution in genetics, ecology and linguistics. *arXiv* cond-mat/0703478v1. 60 pp.

Fukumori K, Livingston G, Leibold MA. 2015. Disturbance-mediated colonization-extinction dynamcis in experimental protist metacommunities. *Ecology* 3234-3242.

Gilpin M. 1994. Community-level competition: asymmetrical dominance. *Proc Natl Acad Sci USA* 91: 3252-3254.

Hubbell SP. 2001. *The unified neutral theory of biodiversity and biogeography*. Princeton University Press, Princeton.

Jessup CM, Forde SE, Bohannan BJM. 2005. Microbial experimental systems in ecology. *Adv Ecol Res* 37: 273-307.

Kimura M. 1953. "Stepping stone" model of population. *Ann Rep Nat Inst Genet Japan* 3: 62–63.

Kimura M, Ohta T. 1971. *Theoretical aspects of population genetics*. Princeton University Press, Princeton.

Leibold MA, Holyoak M, Mouquet N, Amarasekare P, Chase JM, Hoopes MF, Holt RD, Shurin JB, Law R, Tilman D, et al. 2004 The metacommunity concept: a framework for multi-scale community ecology. *Ecol Lett* 7: 601-613.

Leibold MA, Loeuille N. 2015. Species sorting and patch dynamics in harlequin metacommunities affect the relative importance of environment and space. *Ecology* 96: 3227-3233.

Mansour I, Heppell C, Ryo M, Rillig MC. 2018. Application of the microbial community coalescence concept to riverine networks. *Biol Rev* 93: 1832-1845.

Matthiessen B, Hillebrand H. 2006 Dispersal frequency affects local biomass production by controlling local diversity. *Ecol Lett* 9: 652-662.

Mihaljevic JR. 2012. Linking metacommunity theory and symbiont evolutionary ecology. *Trends Ecol Evol* 27: 323-329.

Moran PAP. 1959. The theory of some genetical effects of population subdivision. *Australian J Biol Sci* 12:109-116.

Moran PAP. 1962. *The Statistical Processes of Evolutionary Theory*. Clarendon Press, Oxford.

Parra-Rojas C, McKane AJ. 2018. Reduction of a metapopulation genetic model to an effective one-island model. *arXiv*:1707.07145v2. 22pp.

Perez-Velazquez J, Golgeli M, Garcia-Contreras R. Mathematical modelling of bacterial quorum sensing: a review. *Bull Math Biol* 78: 1585-1639.

Rillig MC, Antonovics J, Caruso T, Lehmann A, Powell JR, Veresoglou SD, Verbruggen E. 2015. Interchange of entire communities: microbial community coalescence. *Trends Ecol Evol* 30: 470-476.

Rillig MC, Lehmann A, Aguilar-Trigueros C, Antonovics J, Caruso T, Hempel S, Lehmann J, Valyi K, Verbruggen E, Veresoglou SD, Powell JR. 2016. Soil microbes and community coalescence. *Pedobiology* 59: 37-40.

Sierocinski P, Milferstedt K, Bayer F, Großkopf T, Alston M, Bastkowski, S, Swarbreck D, Hobbs PJ, Soyer OS, Hamelin J, Buckling A. 2017. The most efficient microbial community dominates during community coalescence. *bioRxiv*, doi: https://doi.org/10.1101/101436.

Rosindell J, Hubbell SP, Etienne RS. 2011. The unified neutral theory of biodiversity and biogeography at age ten. *Trends Ecol Evol* 26: 340-348.

Souza EP, Ferreira EM, Neves AGM. 2018. Fixation probabilities in the Moran process in evolutionary games with two strategies: graph shapes and large population asymptotics. *arXiv*:1805.00117v1. 42pp.

Tikhonov M. 2016. Community-level cohesion without cooperation. *eLife* 5: e15747, 15pp.

Venail PA, MacLean RC, Bouvier T, Brockhurst MA, Hochberg ME, Mouquet N. 2008 Diversity and productivity peak at intermediate dispersal rate in evolving metacommunities. *Nature* 452: 210-214.

Wade MJ, McCauley DE. 1988. Extinction and recolonization: their effects on the genetic differentiation of local populations. *Evolution* 42: 995-1005.

Wakely J, Takahashi T. 2004. The many-demes limit for selection and drift in a subdivided population. *Theor Pop Biol* 66: 83-91.

Wilson DS. 1992 Complex interactions in metacommunities, with implications for biodiversity and higher levels of selection. *Ecology* 73: 1984-2000.

Wright S. 1943. Isolation by distance. *Genetics* 28: 114–156.

# Figure Legends

**Fig. 1.** Time to monodominance in a metacommunity of two species at different levels of subdivision (number of patches) when movement among patches is either by individual dispersal (dotted lines and open dots) or by pairwise coalescence of patches (solid lines and dots). **(a)** Moran model. Individuals moving per time step: red=1, blue=4, black=16. **(b)** Fisher-Wright model. Individuals moving per time step: blue=4, brown=8, black=16. The initial metacommunity consisted of two species each with 64 individuals. In the Fisher-Wright model, the absent points are combinations of movement and subdivision where there was less than one coalescence event per time step. Means are based on 1000 runs, with points being 5 independent runs of 200. Asterisk in lower left shows the theoretical expectation for no subdivision.

**Fig. 2.** Time to monodominance in a metacommunity of two species at different levels of subdivision (number of patches), with coalescence that is either pairwise (dashed lines and solid dots) or diffuse (dotted lines and open dots). **(a)** Moran model. Diffuse coalescence involved 8 patches; to equate net movement by the two types of coalescence, diffuse coalescence occurred at a lower frequency than pairwise coalescence. **(b)** Fisher-Wright model. To equate net movement by the two types of coalescence, diffuse coalescence involved fewer patches than pairwise coalescence (see text). In both models, individuals moving per time step: blue=4, brown=8, black=16. The initial metacommunity consisted of two species each with 64 individuals. In the Fisher-Wright model, the absent points are combinations of movement and subdivision where there was less than one coalescence event per time step. Means are based on 1000 runs, with points being 5 independent runs of 200.

**Fig. 3. (a)** Time to monodominance (log10) for a species with different levels of selective advantage, when there is either dispersal (dotted lines and open dots) or pairwise coalescence (solid lines and dots). Patch number: green= 8, brown =16. **(b)** Probability of favored type achieving dominance at different levels of community subdivision (number of patches) when there is either dispersal (dotted lines and open dots) or pairwise coalescence (solid lines and dots). Selective advantage: black =0.05, blue=0.01, red=0.001. Based on the Moran model of a metacommunity of two species each with 64 individuals. Means are for 5 independent subsets of 200 replicates each.

Fig 1 (a)

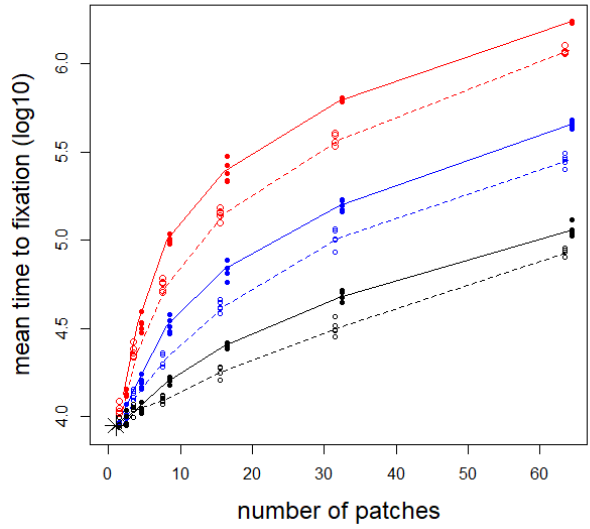

Fig 1 (b)

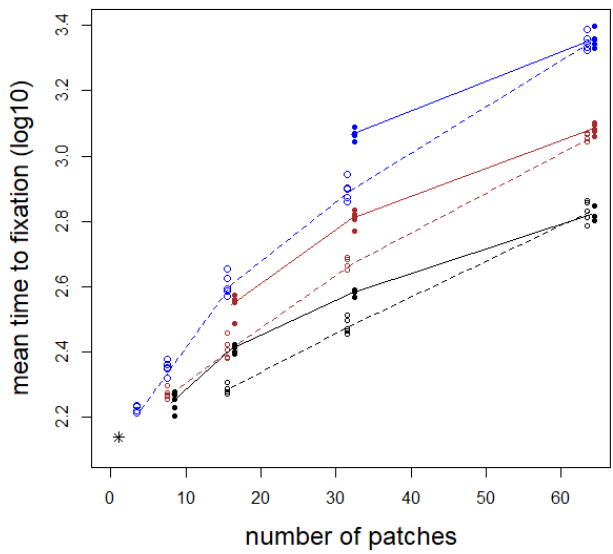

Fig 2 (a)

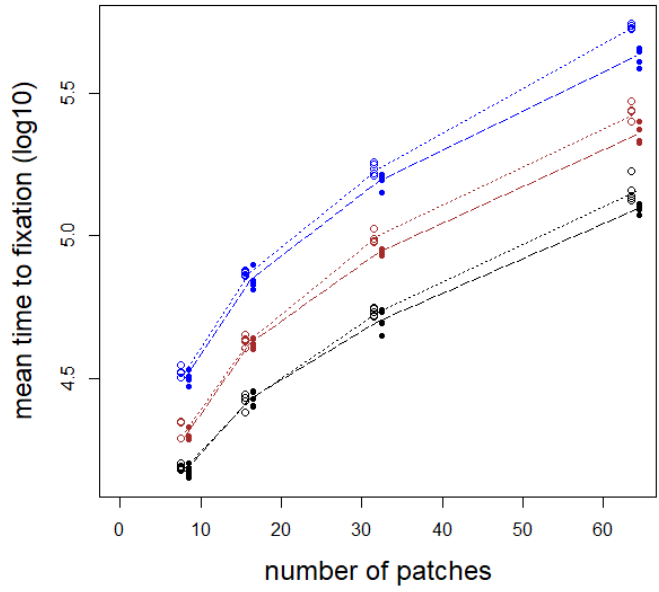

Fig. 2 (b)

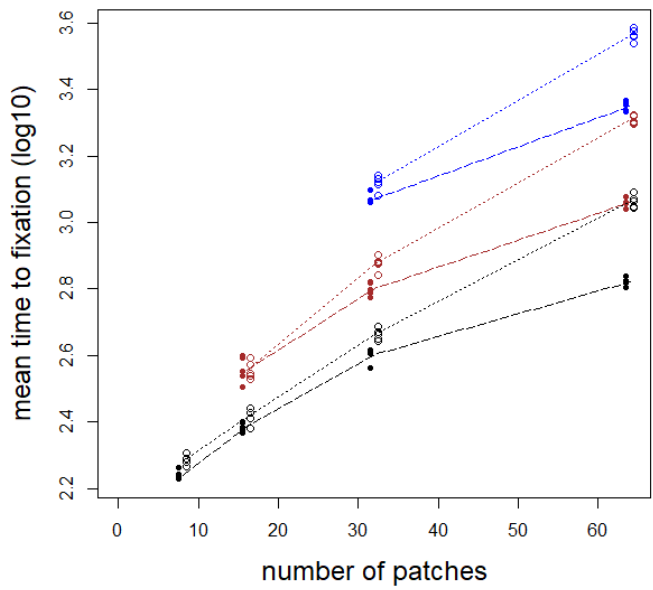

Fig. 3 (a)

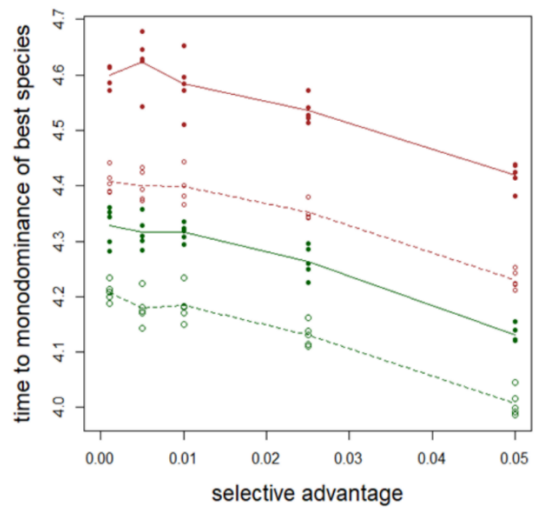

Fig. 3 (b)

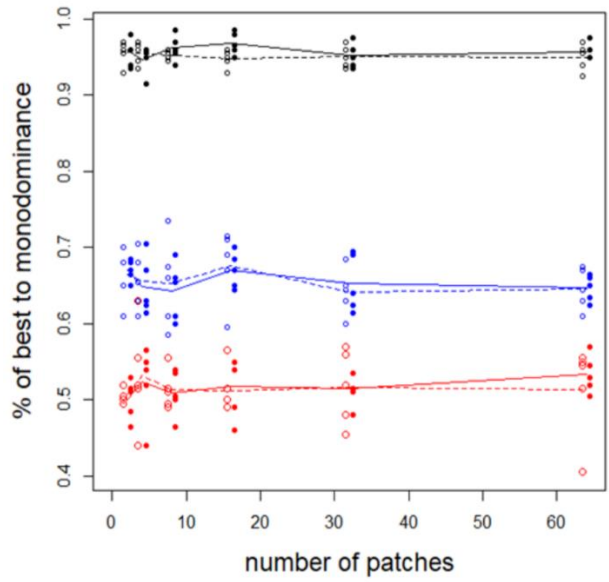

SUPPLEMENTARY MATERIAL

for

**Species diversity in a metacommunity with patches connected by periodic coalescence: a neutral model.**

Janis Antonovics[1], Stavros D. Veresoglou[2,3], Matthias C. Rillig[2,3],

[1] Biology Department, University of Virginia, Charlottesville, VA22904, USA

[2] Institut für Biologie, Freie Universität Berlin, 14195 Berlin, Germany.

[3] Berlin-Brandenburg Institute of Advanced Biodiversity Research (BBIB), 14195 Berlin, Germany.

[1] Corresponding author. Present address: Biology Department, University of Virginia, Charlottesville, VA22904, USA. http://orcid.org/0000-0002-5189-1611

Appendix S1.... **Tables and Figures**

Appendix S2.... **Explanation of Models**

Appendix S3.... **R code used to instantiate the models**

# Appendix S1. Tables and Figures

**Table S1. 1.** Mean and median times to monodominance under a Moran neutral model with dispersal or with paired coalescence. N=128 and total migrants=8 for three levels of patch subdivision, and values are means for 250 replicate runs.

| Number of patches | Mean (back transformed from mean of log time) | | Median | |
|---|---|---|---|---|
| | Dispersal | Coalescence | Dispersal | Coalescence |
| 8 | 22,284.4 | 31,260.8 | 23,107.5 | 35,578.5 |
| 16 | 42,169.7 | 67,920.4 | 41,514.0 | 69,452.0 |
| 32 | 108,143.4 | 169,824.4 | 105,985.5 | 175,049.5 |

**Fig. S1.1.** Mean time to monodominance of a species in a metacommunity under the Fisher-Wright model with deterministic dispersal (dotted lines and open dots) and pairwise coalescence but random assignment of individuals to patches following coalescence. Each point is based on 200 independent runs; standard errors are ca. 2x width of the symbols. Three rates of net movement per time step per metacommunity are shown (red=1 individual; blue=4; black=16). where at each time step all individuals die and are replaced at random from the same patch. Points are absent for combinations of movement and subdivision where dispersal could not be exactly equated with pairwise coalescence within one time-step.

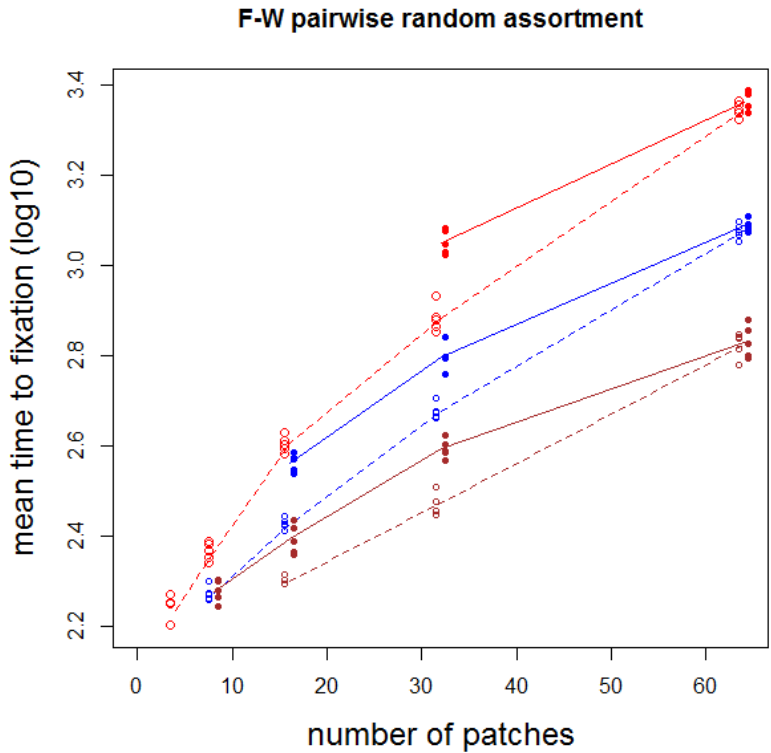

**Fig. S1.2.** Mean time to monodominance of a species in a metacommunity with either pairwise coalescence (dashed lines and open dots) or diffuse coalescence (dotted lines and solid dots) at different levels of community subdivision (number of patches). In this simulation, the same number of patches were involved in diffuse coalescence as in pairwise coalescence (levels of movement were greater with diffuse coalescence). Based on the Fisher-Wright model (see text) and random assignment following coalescence. Means are for 5 independent subsets of 200 replicates each. Four rates of movement for paired coalescence are shown (red=2 individuals per metacommunity of $N$=128; blue=4 brown=8; black=16). Points are absent for combinations of movement and subdivision where there was less than one coalescence event per time step.

Diffuse coalescence (measured by number of patches coalescing) can now result in shorter times to monodominance, especially at high movement rates and intermediate levels of subdivision.

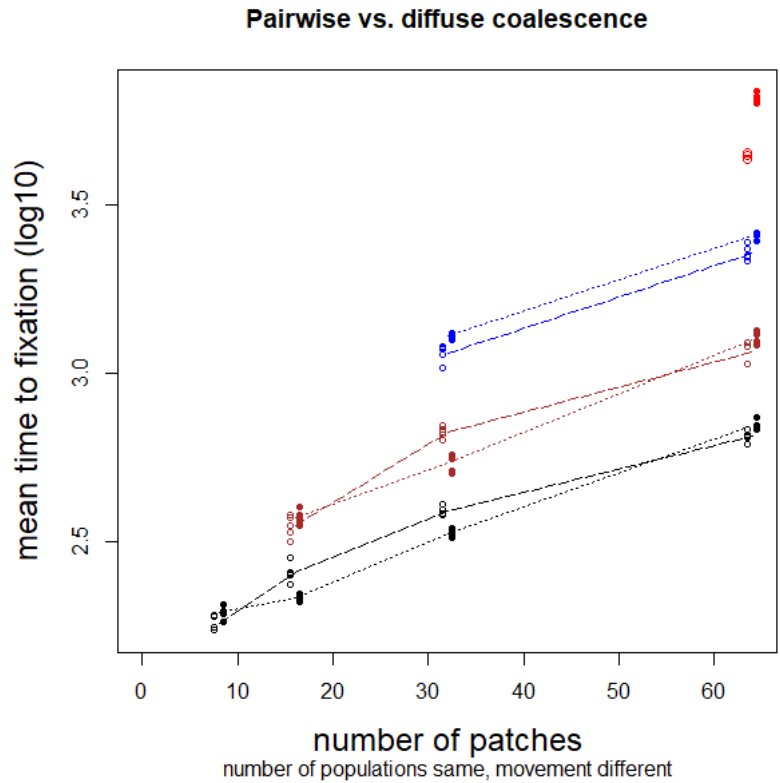

**Fig. S1.3.** Relationship between initial frequency of a species/allele (A) and time to fixation of that species in a community/population. Based on the Moran model (Souza et al. 2018). The Wright-Fisher model gives the same shape curve (Kimura and Ohta 1969, Fig. 1). The intersection of the central vertical line, and the chord of joining the lines on either side shows show that the mean time to fixation of two patches differing in their initial frequencies is less than the time to fixation of two patches with the average of those frequencies.

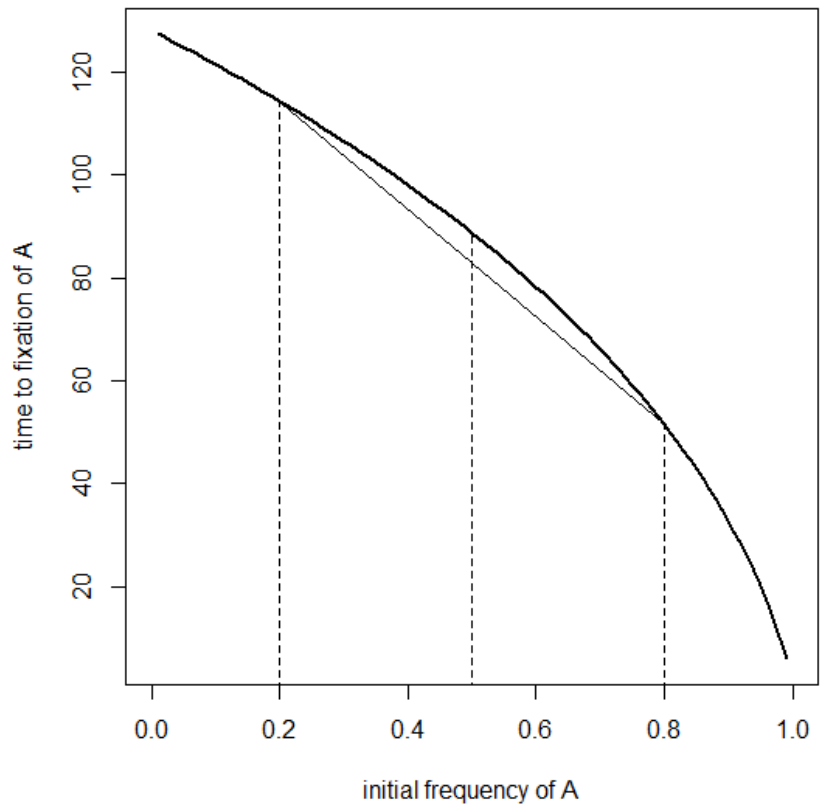

# APPENDIX S2. Explanation of models

## S2.1. Moran and Fisher-Wright models when n=2.

Note – identities (superscripts) are source patches, not individual state

*Dispersal algorithm when n=2, probability = m per time unit*
'Fuse' individuals from two patches $A^1$ and $A^2$, assign one back to each patch at random. The outcome is either $A^1 \mid A^2$ (no dispersals), $A^2 \mid A^1$ (two dispersals); i.e. average = one dispersal/movement.

*Coalescence algorithm when n=2, probability =m/2 per time unit*
'Fuse' two patches $A^1 A^1$ and $A^2 A^2$ and assign two back to each patche at random.
Possible equally likely outcomes are:
  $A^1 A^1 \mid A^2 A^2$ (no moves)
  $A^1 A^2 \mid A^2 A^1$ (2 moves)
  $A^2 A^1 \mid A^1 A^2$ (2 moves)
  $A^2 A^2 \mid A^1 A^1$ (4 moves)
i.e. total of 2 moves every coalescent event but occurs m/n or half as often as disperal when n=2.

If they happen simultaneously (as in the Fisher Wright model) and deterministically then coalescence achieves is the same as dispersal, i.e. dispersal and coalescence produce the same number of net moves simultaneously. However, in the Moran model the movement events happen sequentially and therefore the periods of 'drift' are longer between coalescence events than between dispersal events.
   To confirm this, we predicted that if dispersal/coalescence is made deterministic in the Moran model, there should also be longer times to monodominance, and this was the case (see Table below):
Time to monodominance when $N$ =128 and $p$=64 ($n$=2), with the Moran model when dispersal occurs every generation and coalescence occurs every other generation. Two right hand columns show times to monodominance when dispersal and coalescence are applied probabilistically in the Moran model. Standard errors on means are ca. 0.02-0.03 for 250 replicate runs.

| number of individuals moving | mean log time to monodominance | | median time to monodominance | | mean log time to monodominance in | |
|---|---|---|---|---|---|---|
| | movement deterministic | | | | movement stochastic | |
| | Dispersal | Coalescence | Dispersal | Coalescence | Dispersal | Coalescence |
| 1 | 6.043 | 6.233 | 1,078,645.0 | 1,609,636.0 | 6.044 | 6.205 |
| 8 | 5.426 | 5.594 | 304,594.0 | 410,254.5 | 5.473 | 5.633 |
| 16 | 4.893 | 5.076 | 85,441.5 | 121,501.0 | 4.905 | 5.087 |

## S2.2. Theoretical fixation times

(**from** Kimura and Ohta 1969; and
https://math.la.asu.edu/~jtaylor/teaching/Spring2015/APM504/lectures/Moran.pd
Theorem 3 jetaylor6@asu.edu.)

Assume N haploid individuals, and starting frequency of the alleles are p and 1-p.
Under the Fisher-Wright model the time to fixation is
<u>Either allele:</u> $-2N\{p \log(p) + (1-p) \log(1-p)\}$
<u>One allele:</u>   $-2N\{(1-p)/p) \log(1-p)\}$
For Moran, these Fisher-Wright times are multiplied by N/2 to get predicted times.

Plot of 10000 runs of time to fixation (either allele) in a single population of N=128, Moran model.

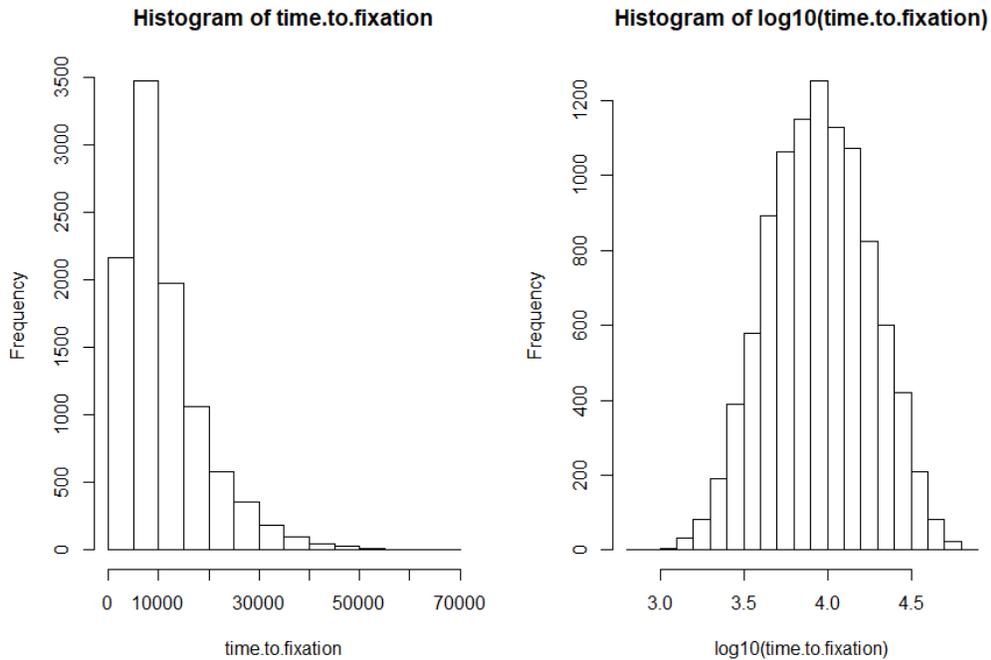

Observed  mean t = 11269.0
Expected  mean t = 11356.5
Mean of log10 t = 3.94936; sterr = 0.00301

Plot of 10000 runs of time to fixation (either allele) in a single population of N=128, Fisher-Wright model.

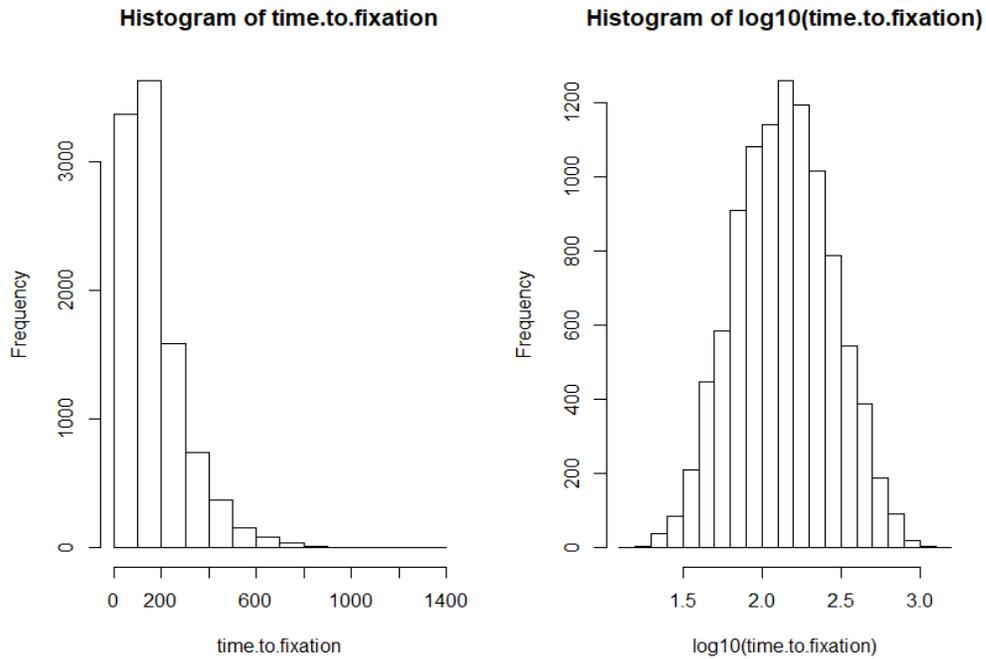

Observed mean t = 174.98
Expected mean t = 177.46
Mean of log10 t = 2.13882; sterr=0.00303

This simulation shows that the log of the mean time to fixation agrees with the theoretical prediction for mean time to fixation.

**S2.3. Number of coalescent events that give equivalent levels of movement as dispersal, for different levels of patch subdivision, illustrated for a community of *N*=128.**

m = movement rate, p =patches, n = individuals within patches

|        |       | m   | 1/128 | 2/128 | 4/128 | 8/128 | 16/128 | 32/128 |
|--------|-------|-----|-------|-------|-------|-------|--------|--------|
| movers | N*p   |     | 1     | 2     | 4     | 8     | 16     | 32     |
| p      | n     |     |       |       |       |       |        |        |
| 2      | 64    |     |       |       |       |       |        |        |
| 4      | 32    |     |       |       |       |       |        | 1      |
| 8      | 16    |     |       |       |       |       | 1      | 2      |
| 16     | 8     |     |       |       |       | 1     | 2      | 4      |
| 32     | 4     |     |       |       | 1     | 2     | 4      | 8      |
| 64     | 2     |     |       | 1     | 2     | 4     | 8      | 16     |

# APPENDIX S3: R code used to instantiate the models

**Note: functions implemented in the code are indicated by names in upper case letters, and the code for them is at the beginning of the code.**

**Moran model**
    **Compare dispersal with pairwise coalescence**
    **Compare pairwise with diffuse coalescence**

**Fisher-Wright model**
    **Compare dispersal with pairwise coalescence**
    **Compare pairwise with diffuse coalescence**

**Moran model: Compare dispersal with pairwise coalescence**

```r
################################################################################
#####
#import packages
library(foreach)
library(doParallel)
cl=makeCluster(7)
registerDoParallel(cl)
######################starting conditions
MO.DISP.PW=function(runs,p,n,m,d,tmax)
        {
 meta.d=matrix(rbinom(p*n,1,0.5),n,p,byrow=F) #generate metapopulation; rows are individuals within pops, columns are pop
 for (t in (1:tmax))
                {
        #moran birth death
                if (runif(1)<d)
                 {
                pid=sample(c(1:p),1,replace=F)   # sample 1 individuals from pop
                nid=sample(c(1:n),1,replace=F)   # sample 1 individual from within pop
                        freq=sum(meta.d[,pid])/n       # calculate frequency in chosen population
                meta.d[nid,pid]=rbinom(1,1,freq) # replace that individual with one chosen from same pop at random
                 }
                #dispersal, instantiated as coalescence with 1; each coalescent event disperses ON AVERAGE 1 individual (half diperse none, half disperse 2)
                if (runif(1)<m) #do if there is a movement event
                 {
        pids=sample(c(1:p),2,replace=F) # sample from different populations in the metapop without replacement
                nids=sample(c(1:n),2,replace=T) # sample individuals from these populations
                n1=nids[1];p1=pids[1]
                n2=nids[2];p2=pids[2]
                coalpop=c(meta.d[n1,p1],meta.d[n2,p2]) # fuse two individuals
                coal=sample(coalpop)                                                     # randomise individuals in fused pops
                meta.d[n1,p1]=coal[1]           # separate them again
                meta.d[n2,p2]=coal[2]
```

```r
                }
            sum.d=sum(meta.d) # find total number of 1's in population
            if (sum.d==0|sum.d==p*n) #if number of 1's are zero or fixed, break
    {break}
            } #end t loop
            hold=c(t,sum.d/N)
  return(hold)
}#end of function MO.DISP.PW
#######################starting conditions
MO.COAL.PW=function(runs,p,n,m,d,tmax)
            {
meta.c=matrix(rbinom(p*n,1,0.5),n,p,byrow=F) #starting metapopulation same in dispersal and coalescence
            for (t in (1:tmax))
            {
                c=m/n
                #moran birth death
                if (runif(1)<d)
                {
            pid=sample(c(1:p),1,replace=F) # sample 1 individuals from metapop, sample pop
            nid=sample(c(1:n),1,replace=T) # sample individual
                freq=sum(meta.c[,pid])/n
            meta.c[nid,pid]=rbinom(1,1,freq) #replace that individual with random from pop
                }
                #coalescence pairwise
            if (runif(1)<c)
                {
            pops=sample(c(1:p),2,replace=F)
                p1=pops[1];p2=pops[2]
                coalpop=c(meta.c[,p1],meta.c[,p2]) # fuse pops
                coal=sample(coalpop)                                          #
randomise individuals in fused pops
                meta.c[,p1]=coal[1:n]          #separate them again - first pop
                meta.c[,p2]=coal[(n+1):(n+n)] #second pop
            }#end condition to do anything
  sum.c=sum(meta.c)
  if (sum.c==0|sum.c==N) #test for fixation
  {
  break
  }
            } #end t loop
  hold=c(t,sum.c/N)
  return(hold)
  } #end of function MO.COAL.PW
  ##################################################################################
SUMSTATS=function(hold,tmax) # summarize over reps
{
pastmax=subset(hold,hold[,1]>=tmax) #array of data that did not go to monodominance
repspastmax=length(pastmax[,1])
fraction1s=sum(hold[,2])/length(hold[,2])
mediant=(median(hold[,1], na.rm=T))    #calculate median
q1=(quantile(hold[,1],c(0.25),na.rm=T)) #calculate quantiles
q3=(quantile(hold[,1],c(0.75),na.rm=T))
meanlt=sum(log10(hold[,1]))/length(hold[,1])       #calculate mean of log times to fixation
```

```r
	stdevlt=sqrt(var(log10(hold[,1])))	#calculate stdev of log times to fixation
	sterrlt=stdevlt/sqrt(length(hold[,1]))		#calculate stderr of log times to fixation
	out=c(meanlt,sterrlt,mediant,q1[[1]],q3[[1]],length(hold[,1]),repspastmax,fraction1s)
	return(out)
} #end SUMSTATS
################################################################################################
####
#end subroutines
################################################################################################
####
overall.start=Sys.time()
################################################################################################
####
N=128
p.step=c(2,4,8,16,32,64)#
m.step=c(4/128,8/128,16/128)# 1/128,
d.step=c(1)
#reps=250
tmax=100000000
superreps=5
############################################################################################
for (i in 1:superreps)
{
hold.all=matrix(NA,2*length(p.step)*length(m.step)*length(d.step),12,byrow=T) #holds final summary data

	# 2 times to accommodate dispesal and coalescence
	
	# Type N p d m meanlt sterrlt meanmedt lower-quartile upper-quartile reps repspastmax fraction-1's
h=1 #counter for rows in the matrix hold

#DISPERSAL (coalescence with 1 individual)
for (p in p.step)
{
n=N/p #get correspoonding pop size given p pops and toal size N
for(m in m.step)
 {
	for(d in d.step)
	 {
	
	######################################################################################
###############
			hold <- foreach(runs=rep(1,20 ), .combine='rbind') %dopar% MO.DISP.PW(runs,p,n,m,d,tmax)
	
	######################################################################################
###############
			rep.output=SUMSTATS(hold,tmax)
			print(c(i,"Dispersal",p,d,m))   #keep track of how long simulation has run
  type=1 # use number 1 for dispersal
			hold.all[h,]=c(type,p,d,m,rep.output)
  h=h+1 #save
	 } # end of d loop
	} # end of m loop
} # end of p loop
```

```r
#COALESCENCE
for (p in p.step)
{
        n=N/p #get correspoonding pop size given p pops and toal size N
 for(m in m.step)
 {
 #c=m/n # coalescence as function of movement (each coalescent event gives n dispersals)
        for(d in d.step)
                {

################################################################################################
                hold <- foreach(runs=rep(1,20 ), .combine='rbind') %dopar% MO.COAL.PW(runs,p,n,m,d,tmax)

        ################################################################################################
                print(c(i,"Coalescence",p,d,m))
                rep.output=SUMSTATS(hold,tmax)
                type=2 # use number not words so output is data.frame
        hold.all[h,]=c(type,p,d,m,rep.output)
  h=h+1 #save
        } # end of d loop
 } # end of m loop
} # end of p loop
r.holdall=nrow(hold.all) #get number of rows in hold.all
name=paste("H",i,sep="") #create super rep names H1, H2, etc (Hi)
assign(name,cbind(matrix(i,r.holdall,1),hold.all)) #assign names to superrep matrices
} # end of superreps loop
super.hold.all= rbind(H1,H2,H3,H4,H5)  #   #MANUALLY ADJUST SUPERREPS if not equal 5
###################################################
overall.time=Sys.time() - overall.start
print(c(overall.time))
################################################   p n sz lam  mov    mean     err median     q1     q3 runs pastmax X1.s
################################################################################################
dat1D=super.hold.all[super.hold.all[,5]==m.step[1] & super.hold.all[,2]==1,]
dat1Dm=aggregate(dat1D[,6], by=list(dat1D[,3]),FUN=mean)#,na.action = na.omit
dat1Dse=sqrt(aggregate(dat1D[,6], by=list(dat1D[,3]),FUN=var))/superreps#,na.action = na.omit
dat1C=super.hold.all[super.hold.all[,5]==m.step[1] & super.hold.all[,2]==2,]
dat1Cm=aggregate(dat1C[,6], by=list(dat1C[,3]),FUN=mean)#,na.action = na.omit
dat1Cse=sqrt(aggregate(dat1C[,6], by=list(dat1C[,3]),FUN=var))/superreps #, na.action = na.omit

dat2D=super.hold.all[super.hold.all[,5]==m.step[2] & super.hold.all[,2]==1,]
dat2Dm=aggregate(dat2D[,6], by=list(dat2D[,3]),FUN=mean)  #,na.action = na.omit
dat2Dse=sqrt(aggregate(dat2D[,6], by=list(dat2D[,3]),FUN=var))/superreps #,na.action = na.omit
dat2C=super.hold.all[super.hold.all[,5]==m.step[2] & super.hold.all[,2]==2,]
dat2Cm=aggregate(dat2C[,6], by=list(dat2C[,3]),FUN=mean)  #,na.action = na.omit
dat2Cse=sqrt(aggregate(dat2C[,6], by=list(dat2C[,3]),FUN=var))/superreps  #,na.action = na.omit

dat3D=super.hold.all[super.hold.all[,5]==m.step[3] & super.hold.all[,2]==1,]
dat3Dm=aggregate(dat3D[,6], by=list(dat3D[,3]),FUN=mean) #,na.action = na.omit
```

```r
dat3Dse=sqrt(aggregate(dat3D[,6], by=list(dat3D[,3]),FUN=var))/superreps  # ,na.action = na.omit
dat3C=super.hold.all[super.hold.all[,5]==m.step[3] & super.hold.all[,2]==2,]
dat3Cm=aggregate(dat3C[,6], by=list(dat3C[,3]),FUN=mean)  #,na.action = na.omit
dat3Cse=sqrt(aggregate(dat3C[,6], by=list(dat3C[,3]),FUN=var))/superreps  #,na.action = na.omit

win.graph()
min=min(super.hold.all[,6],na.rm=T)
max=max(super.hold.all[,6],na.rm=T)
plot(c(0,64),c(min,max),type='n',xlab="number of patches",ylab="mean time to fixation (log10)",cex.lab=1.5,
         main="Moran")

lines(c(dat1Dm[,1]),c(dat1Dm[,2]),lty="dashed",col="red")
#lines(log10(c(dat1Dm[,1])),c(dat1Dm[,2]),lty="dashed",col="red")
points(dat1D[,3]-0.5,dat1D[,6],col="red",cex=1)
lines(dat1Cm[,1],dat1Cm[,2],col="red")
#lines(log10(dat1Cm[,1]),dat1Cm[,2],col="red")
points(dat1C[,3]+0.5,dat1C[,6],col="red",cex=1, pch=20)

lines(dat2Dm[,1],dat2Dm[,2],lty="dashed",col="blue")
points(dat2D[,3]-0.5,dat2D[,6],col="blue",cex=0.75)
lines(dat2Cm[,1],dat2Cm[,2],col="blue")
points(dat2C[,3]+0.5,dat2C[,6],col="blue",cex=1,pch=20)

lines(dat3Dm[,1],dat3Dm[,2],lty="dashed",col="black")
points(dat3D[,3]-0.5,dat3D[,6],col="black",cex=0.75)
lines(dat3Cm[,1],dat3Cm[,2],col="black")
points(dat3C[,3]+0.5,dat3C[,6],col="black",cex=1,pch=20)

points(1,3.949,col="black",cex=1, pch=8) #asterisk=8, plus=3, x=4

#setwd("C:/Users/Janis/Desktop/Coalescence 2019/")
#write.csv(super.hold.all, file="Data Fig 1a.csv")
```

## Fisher-Wright model: Compare dispersal with pairwise coalescence

```r
###########################################################################################
#####
#import packages
library(foreach)
library(doParallel)
cl=makeCluster(7)
registerDoParallel(cl)
######################starting conditions
FW.DISP.DET=function(reps,p,n,m,tmax)
{
 meta.d=matrix(rbinom(p*n,1,0.5),n,p,byrow=F) #generate metapopulation; rows are individuals, columns are pops
 hold=rep(NA,2)
 for (t in (1:tmax))
         {
         #FISHER_WRIGHT BIRTHS-DEATHS
         for (j in (1:p))          #for each population
```

```r
            {
            freq=sum(meta.d[,j])/n      # calculate frequency in population
            meta.d[,j]=rbinom(n,1,freq)   # replace that population with random number of two types based on previous frequency
            }
        #DETERMINISTIC DISPERSAL
  rand.pids=sample(1:p) #randomize populations prior to dispersal
        meta.d=meta.d[,rand.pids]
        d.pops=N*m # different pops involved in dispersal
        if (d.pops>p) #cannot do if to few pops
        {
         sum.d=NA; t=NA # store as NA values
         break
        }
  if (d.pops<=p)
        {
        for (i in (1:(d.pops/2))) #disperse between pairs with no replacement
                {
                n1=sample(c(1:n),1) #identify individuals in 2 different pops
                n2=sample(c(1:n),1)
                temp=meta.d[n1,c((2*i)-1)] # sample an individual from one populations - pops randomized previously
                meta.d[n1,c((2*i)-1)]=meta.d[n2,c(2*i)] # switch with an individual from a different populations
                meta.d[n2,c(2*i)]=temp
        } # e.g. if d.pop=8, this has effected switches between 4 pairs of pops, i.e. N*m=8
        }
  sum.d=sum(meta.d) # find total number of 1's in population
        if (sum.d==0|sum.d==p*n) #if number of 1's are zero or fixed, break
        {
                hold=c(t, sum.d/N)
        break
        }
        } #end t loop

 return(hold)
} # end function MO.DISP.DET
######################starting conditions
FW.COAL.PAIR=function(reps,p,n,m,tmax)
{
 meta.c=matrix(rbinom(p*n,1,0.5),n,p,byrow=F) #starting metapopulation same in dispersal and coalescence
 hold=rep(NA,2)
 for (t in (1:tmax))
        {
        #FISHER-WRIGHT BIRTH-DEATH
        for (j in (1:p))            #for each population
        {
         freq=sum(meta.c[,j])/n      # calculate frequency in population
         meta.c[,j]=rbinom(n,1,freq)   # replace that population with random number of two types based on frequency in previous
        }
        #COALESCENCE PAIRWISE
  rand.pids=sample(1:p)#randomize populations prior to coalescence to simplify selection of pops without replacement
```

```
        meta.c=meta.c[,rand.pids]
        cevents=N*m/n # number of coalescence events
        #e.g. with 1 cevent diffuse coalescencw fuses 2 pops, with 2 fuses 4 pops
        #glob.freq=sum(meta.c)/N
        if (cevents<1)
         {
         sum.c=NA;t=NA
         break
         }
        if (cevents>=1)
        {
        for (i in (1:cevents))
         {

                pop1.half=meta.c[(1:(n/2)),c((2*i)-1)] # sample half from one different populations in the
metapop without replacement
                pop2.half=meta.c[(1:(n/2)),c(2*i)] #sample half from another population
                temp=pop1.half
                meta.c[(1:(n/2)),c((2*i)-1)]=meta.c[1:(n/2),c(2*i)]
                meta.c[(1:(n/2)),c(2*i)]=temp
                meta.c[,c((2*i)-1)]=sample(meta.c[,c((2*i)-1)]) # rerandomize so order is not maintained
                meta.c[,c(2*i)]=sample(meta.c[,c(2*i)])
                } #end cevents
        }#end if cevents
        sum.c=sum(meta.c)
  if (sum.c==0|sum.c==N) #test for fixation
  {
        hold=c(t,sum.c/N)
  break
  }
  } #end t loop

 return(hold)
} #end of function MO.COAL.DIF
############################################################################
SUMSTATS=function(hold,tmax) # summarize over reps
{
 pastmax=subset(hold,hold[,1]>=tmax) #array of data that did not go to monodominance
 repspastmax=length(pastmax[,1])
 fraction1s=sum(hold[,2])/length(hold[,2])
 mediant=(median(hold[,1], na.rm=T))    #calculate median
 q1=(quantile(hold[,1],c(0.25),na.rm=T)) #calculate quantiles
 q3=(quantile(hold[,1],c(0.75),na.rm=T))
 meanlt=sum(log10(hold[,1]))/length(hold[,1])  #calculate mean of log times to fixation
 stdevlt=sqrt(var(log10(hold[,1])))        #calculate stdev of log times to fixation
 sterrlt=stdevlt/sqrt(length(hold[,1]))     #calculate stderr of log times to fixation
 out=c(meanlt,sterrlt,mediant,q1[[1]],q3[[1]],length(hold[,1]),repspastmax,fraction1s)
 return(out)
} #end function SUMSTATS
#############################################################################
####
#end subroutines
```

```r
################################################################################
####
overall.start=Sys.time()
N=128
p.step=c(2,4,8,16,32,64)
m.step=c(4/128,8/128,16/128)#
#reps=1 #dummy variable, assign reps in foreach functions for disp and coal
tmax=10000000
superreps=5 # NB: to change, need also to adjust number of H's below
################################################################################
###
for (i in 1:superreps)
{
hold.all=matrix(NA,2*length(p.step)*length(m.step),11,byrow=T) #holds final summary data for disp and coal
h=1 #line increment for hold.all
#DISPERSAL (coalescence with 1 individual)
for (p in p.step)
{
        n=N/p
 for(m in m.step)
 {

###############################################################################
                hold <- foreach(reps=rep(1,200), .combine='rbind') %dopar% FW.DISP.DET(reps,p,n,m,tmax)

        ###############################################################################
#
                rep.output=SUMSTATS(hold,tmax)# subroutine to do summary stats
                print(c(i,"Dispersal",p,m))   # keep track of how long simulation has run
  type=1 # use number 1 for dispersal
                hold.all[h,]=c(type,p,m,rep.output)
  h=h+1 #save
         } # end of m loop
 } # end of p loop

#COALESCENCE
for (p in p.step)
{
        n=N/p
        for(m in m.step)
        {
 ###############################################################################
         hold <- foreach(reps=rep(1,200), .combine='rbind') %dopar% FW.COAL.PAIR(reps,p,n,m,tmax)

###############################################################################
         rep.output=SUMSTATS(hold,tmax) # subroutine to do summary stats
         print(c(i,"Coalescence",p,m))
         type=2 # use number not words so output is data.frame
         hold.all[h,]=c(type,p,m,rep.output)
  h=h+1 #save
         } # end of m loop
 } # end of p loop
r.holdall=nrow(hold.all) #get number of rows in hold.all
```

```r
 name=paste("H",i,sep="") #create super rep names H1, H2, etc (Hi)
 assign(name,cbind(matrix(i,r.holdall,1),hold.all)) #assign names to superrep matrices
 } # end of superreps loop
super.hold.all= rbind(H1,H2,H3,H4,H5)  #   #MANUALLY ADJUST SUPERREPS if not equal 5
#####################################################
overall.time=Sys.time() - overall.start
print(c(overall.time))
##############################################   p n sz lam   mov     mean       err median    q1
q3 runs pastmax X1.s
##############################################################################################
####################################
dat1D=super.hold.all[super.hold.all[,4]==m.step[1] & super.hold.all[,2]==1,]
dat1Dm=aggregate(dat1D[,5], by=list(dat1D[,3]),FUN=mean)#,na.action = na.omit
dat1Dse=sqrt(aggregate(dat1D[,5], by=list(dat1D[,3]),FUN=var))/superreps#,na.action = na.omit
dat1C=super.hold.all[super.hold.all[,4]==m.step[1] & super.hold.all[,2]==2,]
dat1Cm=aggregate(dat1C[,5], by=list(dat1C[,3]),FUN=mean)#,na.action = na.omit
dat1Cse=sqrt(aggregate(dat1C[,5], by=list(dat1C[,3]),FUN=var))/superreps #, na.action = na.omit

dat2D=super.hold.all[super.hold.all[,4]==m.step[2] & super.hold.all[,2]==1,]
dat2Dm=aggregate(dat2D[,5], by=list(dat2D[,3]),FUN=mean)  #,na.action = na.omit
dat2Dse=sqrt(aggregate(dat2D[,5], by=list(dat2D[,3]),FUN=var))/superreps #,na.action = na.omit
dat2C=super.hold.all[super.hold.all[,4]==m.step[2] & super.hold.all[,2]==2,]
dat2Cm=aggregate(dat2C[,5], by=list(dat2C[,3]),FUN=mean)  #,na.action = na.omit
dat2Cse=sqrt(aggregate(dat2C[,5], by=list(dat2C[,3]),FUN=var))/superreps  #,na.action = na.omit

dat3D=super.hold.all[super.hold.all[,4]==m.step[3] & super.hold.all[,2]==1,]
dat3Dm=aggregate(dat3D[,5], by=list(dat3D[,3]),FUN=mean) #,na.action = na.omit
dat3Dse=sqrt(aggregate(dat3D[,5], by=list(dat3D[,3]),FUN=var))/superreps # ,na.action = na.omit
dat3C=super.hold.all[super.hold.all[,4]==m.step[3] & super.hold.all[,2]==2,]
dat3Cm=aggregate(dat3C[,5], by=list(dat3C[,3]),FUN=mean)  #,na.action = na.omit
dat3Cse=sqrt(aggregate(dat3C[,5], by=list(dat3C[,3]),FUN=var))/superreps  #,na.action = na.omit

win.graph()
min=2.1 #min(super.hold.all[,5],na.rm=T)
max=max(super.hold.all[,5],na.rm=T)
plot(c(0,64),c(min,max),type='n',xlab="number of patches",ylab="mean time to fixation (log10)",cex.lab=1.5)
#,main="F-W Pairwise deterministic"

lines(c(1,dat1Dm[,1]),c(2.14,dat1Dm[,2]),lty="dashed",col="blue")
points(dat1D[,3]-0.5,dat1D[,5],col="blue",cex=1)
lines(dat1Cm[,1],dat1Cm[,2],col="blue")
points(dat1C[,3]+0.5,dat1C[,5],col="blue",cex=1, pch=20)

lines(dat2Dm[,1],dat2Dm[,2],lty="dashed",col="brown")
points(dat2D[,3]-0.5,dat2D[,5],col="brown",cex=0.75)
lines(dat2Cm[,1],dat2Cm[,2],col="brown")
points(dat2C[,3]+0.5,dat2C[,5],col="brown",cex=1,pch=20)

lines(dat3Dm[,1],dat3Dm[,2],lty="dashed",col="black")
points(dat3D[,3]-0.5,dat3D[,5],col="black",cex=0.75)
lines(dat3Cm[,1],dat3Cm[,2],col="black")
points(dat3C[,3]+0.5,dat3C[,5],col="black",cex=1,pch=20)
```

```
##################
points(1,2.139,col="black",cex=1, pch=8)

#setwd("C:/Users/Janis/Desktop/Coalescence 2019/")
#write.csv(super.hold.all, file="Data Fig 1b.csv")
```

## Moran model: Compare pairwise with diffuse coalescence

```
############################################################################################
#####
#import packages
library(foreach)
library(doParallel)
cl=makeCluster(7)
registerDoParallel(cl)
######################starting conditions
MO.COAL.DIFF=function(runs,p,n,m,d,tmax)
        {
 meta.d=matrix(rbinom(p*n,1,0.5),n,p,byrow=F) #generate metapopulation; rows are individuals within pops, columns are pop
 c.d=m/(n*7) #rate of diffuse coalescence
 p.c=8 # number of pops invoivled in coalescence
 for (t in (1:tmax))
                {
        #moran birth death
                #if (runif(1)<d)
                # {
                pid=sample(c(1:p),1,replace=F)   # sample 1 individuals from pop
                nid=sample(c(1:n),1,replace=F)   # sample 1 individual from within pop
                        freq=sum(meta.d[,pid])/n        # calculate frequency in chosen population
                meta.d[nid,pid]=rbinom(1,1,freq) # replace that individual with one chosen from same pop at random
                #}

                if (runif(1)<c.d)
                {
        rand.pids=sample(1:p)#randomize populations prior to coalescence to simplify selection of pops without replacement
        meta.d=meta.d[,rand.pids]
                        # pops involved pairwise = 2*cevents; pops involved diffusely for same movemnt= cevents+1
                select=meta.d[,1:p.c] # choose pops to coalesce - order previously randomized
        vect=as.vector(select)     # turn into a 1-D array
        randa=sample(vect) #randomize individuals in array
        rematrix=matrix(randa,n,p.c,byrow=T)#return to same number of pops
        for (i in 1:p.c)
                {
                 meta.d[,i]=rematrix[,i] #return to same number of pops
         }
                #print(meta.d)
         }#end if c.p events
        sum.d=sum(meta.d) # find total number of 1's in population
```

```r
                if (sum.d==0|sum.d==p*n) #if number of 1's are zero or fixed, break
    {break}
            } #end t loop
            hold=c(t,sum.d/N)
    return(hold)
}#end of function MO.DISP.PW
######################starting conditions
MO.COAL.PW=function(runs,p,n,m,d,tmax)
            {
 meta.c=matrix(rbinom(p*n,1,0.5),n,p,byrow=F) #starting metapopulation same in dispersal and coalescence
            c=m/n #rate of pairwise coalescence
 for (t in (1:tmax))
            {
                    #moran birth death
                        #if (runif(1)<d)
                    #{
            pid=sample(c(1:p),1,replace=F) # sample 1 individuals from metapop, sample pop
            nid=sample(c(1:n),1,replace=T) # sample individual
                    freq=sum(meta.c[,pid])/n
            meta.c[nid,pid]=rbinom(1,1,freq) #replace that individual with random from pop
                        #}

                    #coalescence pairwise

                    if (runif(1)<c)
                    {
                            rand.pids=sample(1:p)#randomize populations prior to coalescence to simplify selection of pops without replacement
            meta.c=meta.c[,rand.pids]

             pops=sample(c(1:p),2,replace=F)
                    p1=pops[1];p2=pops[2]
                    coalpop=c(meta.c[,p1],meta.c[,p2]) # fuse pops
                    coal=sample(coalpop)                                             # randomise individuals in fused pops
                    meta.c[,p1]=coal[1:n]          #separate them again - first pop
                    meta.c[,p2]=coal[(n+1):(n+n)] #second pop
            }#end condition to do anything
 sum.c=sum(meta.c)
 if (sum.c==0|sum.c==N) #test for fixation
  {
  break
  }
        } #end t loop
 hold=c(t,sum.c/N)
 return(hold)
} #end of function MO.COAL.PW
################################################################################
SUMSTATS=function(hold,tmax) # summarize over reps
{
 pastmax=subset(hold,hold[,1]>=tmax) #array of data that did not go to monodominance
 repspastmax=length(pastmax[,1])
 fraction1s=sum(hold[,2])/length(hold[,2])
```

```
mediant=(median(hold[,1], na.rm=T))     #calculate median
q1=(quantile(hold[,1],c(0.25),na.rm=T)) #calculate quantiles
q3=(quantile(hold[,1],c(0.75),na.rm=T))
meanlt=sum(log10(hold[,1]))/length(hold[,1])        #calculate mean of log times to fixation
stdevlt=sqrt(var(log10(hold[,1])))      #calculate stdev of log times to fixation
sterrlt=stdevlt/sqrt(length(hold[,1]))           #calculate stderr of log times to fixation
out=c(meanlt,sterrlt,mediant,q1[[1]],q3[[1]],length(hold[,1]),repspastmax,fraction1s)
return(out)
} #end SUMSTATS
################################################################################
####
#end subroutines
################################################################################
####
overall.start=Sys.time()
################################################################################
####
N=128
p.step=c(8,16,32,64) #2,4,
m.step=c(4/128,8/128,16/128)
d.step=c(1)
#reps=250
tmax=100000000
superreps=5
##############################################################################
for (i in 1:superreps)
{
hold.all=matrix(NA,2*length(p.step)*length(m.step)*length(d.step),12,byrow=T) #holds final summary data

         # 2 times to accommodate dispesal and coalescence

         # Type N p d m meanlt sterrlt meanmedt lower-quartile upper-quartile reps repspastmax fraction-1's
h=1 #counter for rows in the matrix hold

#DISPERSAL (coalescence with 1 individual)
for (p in p.step)
{
n=N/p #get correspoonding pop size given p pops and toal size N
for(m in m.step)
 {
         for(d in d.step)
          {

         ################################################################################
##############
                 hold <- foreach(runs=rep(1,200 ), .combine='rbind') %dopar% MO.COAL.DIFF(runs,p,n,m,d,tmax)

         ################################################################################
###############
                 rep.output=SUMSTATS(hold,tmax)
                 print(c(i,"Dispersal",p,d,m))    #keep track of how long simulation has run
  type=1 # use number 1 for dispersal
                 hold.all[h,]=c(type,p,d,m,rep.output)
```

```r
  h=h+1 #save
                } # end of d loop
        } # end of m loop
} # end of p loop

#COALESCENCE
for (p in p.step)
{
        n=N/p #get correspoonding pop size given p pops and toal size N
 for(m in m.step)
 {
  #c=m/n # coalescence as function of movement (each coalescent event gives n dispersals)
        for(d in d.step)
                {

################################################################################################
                hold <- foreach(runs=rep(1,200 ), .combine='rbind') %dopar% MO.COAL.PW(runs,p,n,m,d,tmax)

        ################################################################################################
                print(c(i,"Coalescence",p,d,m))
                rep.output=SUMSTATS(hold,tmax)
                type=2 # use number not words so output is data.frame
        hold.all[h,]=c(type,p,d,m,rep.output)
  h=h+1 #save
        } # end of d loop
 } # end of m loop
} # end of p loop
 r.holdall=nrow(hold.all) #get number of rows in hold.all
 name=paste("H",i,sep="") #create super rep names H1, H2, etc (Hi)
 assign(name,cbind(matrix(i,r.holdall,1),hold.all)) #assign names to superrep matrices
} # end of superreps loop
super.hold.all= rbind(H1,H2,H3,H4,H5)  #  #MANUALLY ADJUST SUPERREPS if not equal 5

####################################################
overall.time=Sys.time() - overall.start
print(c(overall.time))
##################################################   p n sz lam  mov    mean    err median   q1
q3 runs pastmax X1.s
################################################################################################
####################################
dat1D=super.hold.all[super.hold.all[,5]==m.step[1] & super.hold.all[,2]==1,]
dat1Dm=aggregate(dat1D[,6], by=list(dat1D[,3]),FUN=mean)#,na.action = na.omit
dat1Dse=sqrt(aggregate(dat1D[,6], by=list(dat1D[,3]),FUN=var))/superreps#,na.action = na.omit
dat1C=super.hold.all[super.hold.all[,5]==m.step[1] & super.hold.all[,2]==2,]
dat1Cm=aggregate(dat1C[,6], by=list(dat1C[,3]),FUN=mean)#,na.action = na.omit
dat1Cse=sqrt(aggregate(dat1C[,6], by=list(dat1C[,3]),FUN=var))/superreps #, na.action = na.omit

dat2D=super.hold.all[super.hold.all[,5]==m.step[2] & super.hold.all[,2]==1,]
dat2Dm=aggregate(dat2D[,6], by=list(dat2D[,3]),FUN=mean)  #,na.action = na.omit
```

```r
dat2Dse=sqrt(aggregate(dat2D[,6], by=list(dat2D[,3]),FUN=var))/superreps #,na.action = na.omit
dat2C=super.hold.all[super.hold.all[,5]==m.step[2] & super.hold.all[,2]==2,]
dat2Cm=aggregate(dat2C[,6], by=list(dat2C[,3]),FUN=mean)  #,na.action = na.omit
dat2Cse=sqrt(aggregate(dat2C[,6], by=list(dat2C[,3]),FUN=var))/superreps #,na.action = na.omit

dat3D=super.hold.all[super.hold.all[,5]==m.step[3] & super.hold.all[,2]==1,]
dat3Dm=aggregate(dat3D[,6], by=list(dat3D[,3]),FUN=mean) #,na.action = na.omit
dat3Dse=sqrt(aggregate(dat3D[,6], by=list(dat3D[,3]),FUN=var))/superreps  # ,na.action = na.omit
dat3C=super.hold.all[super.hold.all[,5]==m.step[3] & super.hold.all[,2]==2,]
dat3Cm=aggregate(dat3C[,6], by=list(dat3C[,3]),FUN=mean)  #,na.action = na.omit
dat3Cse=sqrt(aggregate(dat3C[,6], by=list(dat3C[,3]),FUN=var))/superreps  #,na.action = na.omit

win.graph()
min=min(super.hold.all[,6],na.rm=T)
max=max(super.hold.all[,6],na.rm=T)
plot(c(0,64),c(min,max),type='n',xlab="number of patches",ylab="mean time to fixation (log10)",cex.lab=1.5)# main="Moran"

lines(c(dat1Dm[,1]),c(dat1Dm[,2]),lty="dotted",col="blue")
points(dat1D[,3]-0.5,dat1D[,6],col="blue",cex=1)
lines(dat1Cm[,1],dat1Cm[,2],col="blue",lty="longdash")
points(dat1C[,3]+0.5,dat1C[,6],col="blue",cex=1, pch=20)

lines(dat2Dm[,1],dat2Dm[,2],lty="dotted",col="brown")
points(dat2D[,3]-0.5,dat2D[,6],col="brown",cex=1)
lines(dat2Cm[,1],dat2Cm[,2],col="brown",lty="longdash")
points(dat2C[,3]+0.5,dat2C[,6],col="brown",cex=1,pch=20)

lines(dat3Dm[,1],dat3Dm[,2],lty="dotted",col="black")
points(dat3D[,3]-0.5,dat3D[,6],col="black",cex=1)
lines(dat3Cm[,1],dat3Cm[,2],col="black",lty="longdash")
points(dat3C[,3]+0.5,dat3C[,6],col="black",cex=1,pch=20)

#points(1,3.949,col="black",cex=1, pch=8) #asterisk=8, plus=3, x=4

#setwd("C:/Users/Janis/Desktop/")
#write.csv(super.hold.all, file="Data Fig 2a.csv")
```

**Fisher-Wright model: Compare pairwise with diffuse coalescence**

```r
################################################################################
#####
#import packages
library(foreach)
library(doParallel)
cl=makeCluster(7)
registerDoParallel(cl)
######################starting conditions
FW.COAL.PAIR=function(reps,p,n,m,tmax)
 {
  meta.c=matrix(rbinom(p*n,1,0.5),n,p,byrow=F) #starting metapopulation same in dispersal and coalescence
  for (t in (1:tmax))
```

```r
            {
            #FISHER-WRIGHT BIRTH-DEATH
            for (j in (1:p))           #for each population
             {
             freq=sum(meta.c[,j])/n       # calculate frequency in population
              meta.c[,j]=rbinom(n,1,freq)    # replace that population with random number of two types based on
frequency in previous
             }
            #COALESCENCE PAIRWISE
  rand.pids=sample(1:p)#randomize populations prior to coalescence to simplify selection of pops without
replacement
            meta.c=meta.c[,rand.pids]
            cevents=N*m/n # number of coalescence events
            #e.g. with 1 cevent diffuse coalescencw fuses 2 pops, with 2 fuses 4 pops
            #glob.freq=sum(meta.c)/N
            if (cevents<1)
             {
             sum.c=NA;t=NA
             break
             }
            if (cevents>=1)
            {
            for (i in (1:cevents))
             {
             pop1.half=meta.c[(1:(n/2)),i] # sample half from one different populations in the metapop without
replacement
                    pop2.half=meta.c[(1:(n/2)),(cevents+i)] #sample half from another population
                    temp=pop1.half
                    meta.c[(1:(n/2)),i]=meta.c[1:(n/2),(cevents+i)]
                    meta.c[(1:(n/2)),(cevents+i)]=temp
                    meta.c[,i]=sample(meta.c[,i]) # rerandomize so order is not maintained
                    meta.c[,(cevents+i)]=sample(meta.c[,(cevents+i)])
            } #end cevents
            }#end if cevents
            sum.c=sum(meta.c)
 if (sum.c==0|sum.c==N) #test for fixation
  {
  break
  }
 } #end t loop
 hold=c(t,sum.c/N)
 return(hold)
} #
##################################
FW.COAL.DIF=function(reps,p,n,m,tmax)
{
 meta.c=matrix(rbinom(p*n,1,0.5),n,p,byrow=F) #starting metapopulation same in dispersal and coalescence
 for (t in (1:tmax))
            {
            #FISHER-WRIGHT BIRTH-DEATH
            for (j in (1:p))           #for each population
             {
             freq=sum(meta.c[,j])/n       # calculate frequency in population
```

```
                meta.c[,j]=rbinom(n,1,freq)    # replace that population with random number of two types based on frequency in previous
            }
            #COALESCENCE DIFFUSE
   rand.pids=sample(1:p)#randomize populations prior to coalescence to simplify selection of pops without replacement
   meta.c=meta.c[,rand.pids]
            cevents=N*m/n # number of coalescence events
            pops=N*m/n*(p-1) #e.g. with 1 cevent diffuse coalescencw fuses 2 pops, with 2 fuses 4 pops
            #glob.freq=sum(meta.c)/N
            if (cevents<1)
             {
             sum.c=NA;t=NA
             break
             }
            if (cevents>=1) # pops involved in pairwise pp = 2*c; which is (pp+2)/2
            {
            select=meta.c[,1:c(cevents+1)] # choose pops to coalesce - order previously randomized
            vect=as.vector(select)     # turn into a 1-D array
            randa=sample(vect) #randomize individuals in array
            rematrix=matrix(randa,n,c(cevents+1),byrow=T)#return to same number of pops
            for (i in 1:c(cevents+1))
                    {
                    meta.c[,i]=rematrix[,i] #return to same number of pops
             }
            sum.c=sum(meta.c)
            }#end if cevents
   if (sum.c==0|sum.c==N) #test for fixation
    {
    break
    }
  } #end t loop
 hold=c(t,sum.c/N)
 return(hold)
} #end of function MO.COAL.DIF
####################################################################################
SUMSTATS=function(hold,tmax) # summarize over reps
{
pastmax=subset(hold,hold[,1]>=tmax) #array of data that did not go to monodominance
repspastmax=length(pastmax[,1])
fraction1s=sum(hold[,2])/length(hold[,2])
mediant=(median(hold[,1], na.rm=T))    #calculate median
q1=(quantile(hold[,1],c(0.25),na.rm=T)) #calculate quantiles
q3=(quantile(hold[,1],c(0.75),na.rm=T))
meanlt=sum(log10(hold[,1]))/length(hold[,1])  #calculate mean of log times to fixation
stdevlt=sqrt(var(log10(hold[,1])))         #calculate stdev of log times to fixation
sterrlt=stdevlt/sqrt(length(hold[,1]))      #calculate stderr of log times to fixation
out=c(meanlt,sterrlt,mediant,q1[[1]],q3[[1]],length(hold[,1]),repspastmax,fraction1s)
return(out)
} #end function SUMSTATS
####################################################################################
####
#end subroutines
```

```r
################################################################################
####
overall.start=Sys.time()
N=128
p.step=c(2,4,8,16,32,64)
m.step=c(4/128,8/128,16/128) #2/128,
#reps=1 #dummy variable, assign reps in foreach functions for disp and coal
tmax=10000000
superreps=5 # NB: to change, need also to adjust number of H's below
################################################################################
###
for (i in 1:superreps)
 {
hold.all=matrix(NA,2*length(p.step)*length(m.step),11,byrow=T) #holds final summary data for disp and coal
h=1 #line increment for hold.all
#PAIRWISE
for (p in p.step)
 {
        n=N/p
for(m in m.step)
  {

###########################################################################
                hold <- foreach(reps=rep(1,200), .combine='rbind') %dopar% FW.COAL.PAIR(reps,p,n,m,tmax)

        ###########################################################################
#
                rep.output=SUMSTATS(hold,tmax)# subroutine to do summary stats
                print(c(i,"Pairwise",p,m))   # keep track of how long simulation has run
  type=1 # use number 1 for pairwise coalescence
                hold.all[h,]=c(type,p,m,rep.output)
  h=h+1 #save
          } # end of m loop
 } # end of p loop

#DIFFUSE
for (p in p.step)
 {
         n=N/p
        for(m in m.step)
          {
 ###########################################################################
         hold <- foreach(reps=rep(1,200), .combine='rbind') %dopar% FW.COAL.DIF(reps,p,n,m,tmax)

###########################################################################
        rep.output=SUMSTATS(hold,tmax) # subroutine to do summary stats
        print(c(i,"Diffuse",p,m))
        type=2 # use number 2 for diffuse coalescence
        hold.all[h,]=c(type,p,m,rep.output)
  h=h+1 #save
          } # end of m loop
 } # end of p loop
r.holdall=nrow(hold.all) #get number of rows in hold.all
```

```r
 name=paste("H",i,sep="") #create super rep names H1, H2, etc (Hi)
 assign(name,cbind(matrix(i,r.holdall,1),hold.all)) #assign names to superrep matrices
 } # end of superreps loop
super.hold.all= rbind(H1,H2,H3,H4,H5)  # ,H5  #MANUALLY ADJUST SUPERREPS if not equal 5
########################################################
overall.time=Sys.time() - overall.start
print(c(overall.time))
##############################################   p n sz lam  mov    mean     err median    q1
q3 runs pastmax X1.s
#################################################################################################
#####################################
dat1D=super.hold.all[super.hold.all[,4]==m.step[1] & super.hold.all[,2]==1,]
dat1Dm=aggregate(dat1D[,5], by=list(dat1D[,3]),FUN=mean)#,na.action = na.omit
dat1Dse=sqrt(aggregate(dat1D[,5], by=list(dat1D[,3]),FUN=var))/superreps#,na.action = na.omit

dat1C=super.hold.all[super.hold.all[,4]==m.step[1] & super.hold.all[,2]==2,]
dat1Cm=aggregate(dat1C[,5], by=list(dat1C[,3]),FUN=mean)#,na.action = na.omit
dat1Cse=sqrt(aggregate(dat1C[,5], by=list(dat1C[,3]),FUN=var))/superreps #, na.action = na.omit

dat2D=super.hold.all[super.hold.all[,4]==m.step[2] & super.hold.all[,2]==1,]
dat2Dm=aggregate(dat2D[,5], by=list(dat2D[,3]),FUN=mean)  #,na.action = na.omit
dat2Dse=sqrt(aggregate(dat2D[,5], by=list(dat2D[,3]),FUN=var))/superreps #,na.action = na.omit
dat2C=super.hold.all[super.hold.all[,4]==m.step[2] & super.hold.all[,2]==2,]
dat2Cm=aggregate(dat2C[,5], by=list(dat2C[,3]),FUN=mean)  #,na.action = na.omit
dat2Cse=sqrt(aggregate(dat2C[,5], by=list(dat2C[,3]),FUN=var))/superreps  #,na.action = na.omit

dat3D=super.hold.all[super.hold.all[,4]==m.step[3] & super.hold.all[,2]==1,]
dat3Dm=aggregate(dat3D[,5], by=list(dat3D[,3]),FUN=mean) #,na.action = na.omit
dat3Dse=sqrt(aggregate(dat3D[,5], by=list(dat3D[,3]),FUN=var))/superreps  # ,na.action = na.omit
dat3C=super.hold.all[super.hold.all[,4]==m.step[3] & super.hold.all[,2]==2,]
dat3Cm=aggregate(dat3C[,5], by=list(dat3C[,3]),FUN=mean)  #,na.action = na.omit
dat3Cse=sqrt(aggregate(dat3C[,5], by=list(dat3C[,3]),FUN=var))/superreps  #,na.action = na.omit

#dat4D=super.hold.all[super.hold.all[,4]==m.step[4] & super.hold.all[,2]==1,]
#dat4Dm=aggregate(dat4D[,5], by=list(dat4D[,3]),FUN=mean)  #,na.action = na.omit
#dat4Dse=sqrt(aggregate(dat4D[,5], by=list(dat4D[,3]),FUN=var))/superreps #,na.action = na.omit
#dat4C=super.hold.all[super.hold.all[,4]==m.step[4] & super.hold.all[,2]==2,]
#dat4Cm=aggregate(dat4C[,5], by=list(dat4C[,3]),FUN=mean)  #,na.action = na.omit
#dat4Cse=sqrt(aggregate(dat4C[,5], by=list(dat4C[,3]),FUN=var))/superreps  #,na.action = na.omit

win.graph()
min=min(super.hold.all[,5],na.rm=T)
max=max(super.hold.all[,5],na.rm=T)
plot(c(0,64),c(min,max),type='n',xlab="number of patches",ylab="mean time to fixation (log10)",cex.lab=1.5)
         # ,main= "Pairwise vs. diffuse coalescence",sub="standardized for equal movement")

lines(dat1Dm[,1],dat1Dm[,2],lty="longdash",col="blue")
points(dat1D[,3]-0.5,dat1D[,5],col="blue",cex=1, pch=20)
lines(dat1Cm[,1],dat1Cm[,2],col="blue",lty="dotted")
points(dat1C[,3]+0.5,dat1C[,5],col="blue",cex=1)

lines(dat2Dm[,1],dat2Dm[,2],lty="longdash",col="brown")
points(dat2D[,3]-0.5,dat2D[,5],col="brown",cex=1, pch=20)
```

```
lines(dat2Cm[,1],dat2Cm[,2],col="brown",lty="dotted")
points(dat2C[,3]+0.5,dat2C[,5],col="brown",cex=1)

lines(dat3Dm[,1],dat3Dm[,2],lty="longdash",col="black")
points(dat3D[,3]-0.5,dat3D[,5],col="black",cex=1, pch=20)
lines(dat3Cm[,1],dat3Cm[,2],col="black",lty="dotted")
points(dat3C[,3]+0.5,dat3C[,5],col="black",cex=1)

#lines(dat4Dm[,1],dat4Dm[,2],lty="longdash",col="black")
#points(dat4D[,3]-0.5,dat4D[,5],col="black",cex=0.75)
#lines(dat4Cm[,1],dat4Cm[,2],col="black",lty="dotted")
#points(dat4C[,3]+0.5,dat4C[,5],col="black",cex=1,pch=20)

#setwd("C:/Users/Janis Antonovics/Desktop/")
#write.csv(super.hold.all, file="Data Fig 2b.csv")
```